\documentclass[%
 reprint,
 amsmath,amssymb,
 aps,
pra,
]{revtex4-2}

\usepackage{xcolor}
\usepackage{graphicx}
\usepackage{dcolumn}
\usepackage{makecell}
\usepackage{multirow}
\usepackage{hyperref}
\usepackage[normalem]{ulem}

\hypersetup{
    colorlinks = true,
    citecolor = blue,
    linkcolor = blue,
    filecolor = magenta,      
    urlcolor = .,
    }

\makeatletter
\newsavebox{\@brx}
\newcommand{\llangle}[1][]{\savebox{\@brx}{$\m@th{#1\langle}$}%
  \mathopen{\copy\@brx\kern-0.5\wd\@brx\usebox{\@brx}}}
\newcommand{\rrangle}[1][]{\savebox{\@brx}{$\m@th{#1\rangle}$}%
  \mathclose{\copy\@brx\kern-0.5\wd\@brx\usebox{\@brx}}}
\makeatother

\begin{document}

\preprint{APS/123-QED}

\title{Global Entanglement Quantification via Classical Shadows}

\author{João P. Engster}
    \email{engsterjp@gmail.com}
    \affiliation{Departamento de Física, Universidade Federal de Santa Catarina, Campus Trindade, Florianópolis 88040-900, SC,
Brazil}

\author{Eduardo I. Duzzioni}
    \email{duzzioni@gmail.com}
    \affiliation{Departamento de Física, Universidade Federal de Santa Catarina, Campus Trindade, Florianópolis 88040-900, SC,
Brazil}




\date{\today}

\begin{abstract}
Detecting and characterizing quantum correlations are tasks of great relevance in quantum information. More specifically, quantifying the amount of multipartite entanglement is a known difficult task, even for pure states. To this end, several entanglement measures have been proposed, although there is currently no universal manner to do so. In this work, we propose the classical shadows technique to measure the generalized global entanglement introduced in Phys. Rev. A 74, 022314. This particular multipartite entanglement quantifier $E_G^{(n)}$ relies on the linear entropies of all $n$-qubit partitions of a state. As the linear entropy can be written in terms of a complete set of observables of the subsystem, we employ classical shadows to estimate many observables with fewer measurements. We simulate the quantification of $E_G^{(n)}$ for well-known entangled states and for random states, comparing shadow estimations and grouping techniques. Our results show a clear advantage of classical shadows over direct estimation approaches, indicating that it can be a useful tool to quantify entanglement without the need to reconstruct the density operator of the whole system.
\end{abstract}

\maketitle


\section{Introduction}
\label{sec: introduction}

The characterization of entanglement in quantum systems is a pivotal point in many protocols and tasks in quantum information processing, such as quantum teleportation~\cite{Ren_2017, Bennett_1993}, quantum cryptography, and quantum key distribution~\cite{Jennewein_2000, Ekert_1991, Durt_2003}. In quantum computing, entanglement is one of the key ingredients not only for realizing quantum algorithms, but for achieving speedups~\cite{Jozsa_2003, Jozsa_1997}. In this manner, quantum hardware across different platforms has recently demonstrated entanglement involving an increasing number of qubits~\cite{Monz_2011, Song_2017, Wang_2018, Gong_2019}, reassuring the need to precisely quantify the amount of entanglement in many-body quantum systems. However, there are many different ways of doing so, and with different entanglement measures come different issues and benefits.

A proper entanglement measure $E(\rho)$ should possess some basic properties~\cite{Guhne_2009, Bennett_1996}. It should vanish for separable states, be invariant for local changes of bases, and not increase under local operations and classical communication (LOCC), given that LOCC cannot create entanglement. Additionally, $E(\rho)$ should be convex, meaning that it should decrease under the mixing of states. For instance, the von Neumann entanglement entropy is a common measure, 
\begin{equation}
    S(\rho) = -\text{Tr} \left( \rho \ln \rho \right),
    \label{eq: von neumann entropy}
\end{equation}
since for a pure bipartite state $\rho = |\psi\rangle_{AB}\langle \psi |$, separability implies $S(\rho_A) =S(\rho_B)= 0$. The case for multipartite entanglement is more intricate, and measures to quantify multipartite entanglement often rely on optimizations over states~\cite{Ma_2024}, which are in general not practical to compute~\cite{Weinbrenner_2025}. To address this issue, some methods have been proposed to assess geometric measures of multipartite entanglement~\cite{Wei_2003, Sen_2010, Blasone_2008, Cianciaruso_2016}. In particular, the generalized global entanglement measure $E_G^{(n)}$, introduced in Ref.~\cite{Rigolin_2006}, enables the quantification of multipartite entanglement of pure states, at the cost of estimating a large number of observables (Pauli strings), which can grow rapidly depending on the size of the subsystems. From the expected values of such observables, the entanglement is then evaluated by averaging the linear entropies of all possible $n$-qubit partitions of the quantum state. However, tasks involving the estimation of many expected values suffer from scaling issues: as the system size grows, accurately estimating a large number of observables poses a significant challenge for experimental purposes, demanding an impractical measurement budget, as is the case for quantum chemistry~\cite{Gonthier_2022}. To tackle the feasibility of this measure from a more realistic point of view, we study $E_G^{(n)}$ focusing on the measurement process. In particular, we compare different methods to estimate the Pauli strings, employing classical shadows (CS)~\cite{Huang_2020}, qubit-wise commutativity (QWC)~\cite{Verteletskyi_2020_qwc}, and full commutativity (FC)~\cite{Yen2020}.

The manuscript is organized as follows: in Sec.~\ref{sec: materials} we discuss the main tools used in our study. We begin presenting the generalized global entanglement measure, detailing how it quantifies entanglement operationally by the estimation of Pauli observables. Then, we review the protocol of classical shadows, with particular emphasis on how this technique is applied to observable estimation and how it can be useful to tackle $E_G^{(n)}$. We wrap Sec.~\ref{sec: materials} with a brief discussion of direct estimation methods based on grouping compatible observables to optimize the use of experimental resources. Following up, in Sec.~\ref{sec: results}, we discuss our results, comparing the methods previously mentioned for GHZ, W, EPR, and random states. Lastly, we outline our conclusions and perspectives in Sec.~\ref{sec: conclusion}.
\section{Methods}
\label{sec: materials}

\subsection{Generalized global entanglement}
\label{subsec: generalized entanglement}

The notion of global entanglement, introduced in Ref.~\cite{Meyer_2002}, builds upon the purities of single-qubit reduced density matrices, and leads to a quantifier of entanglement given simply by the average of their linear entropies~\cite{Barnum_2004, Brennen_2003, Lakshminarayan_2005}. The generalization of global entanglement, denoted by $E_G^{(n)}$~\cite{Oliveira_2006, Rigolin_2006}, is then based on the linear entropies of all possible $n$-qubit partitions of the state. It provides an operational approach to quantify the amount of entanglement in a multi qubit system, i.e., it is easily computable for qubit systems, at the cost of estimating a many $n$-local Pauli operators. This measure has been applied in the study of many physical systems of interest, giving insights in quantum phase transitions and genuine multipartite entanglement~\cite{de_Oliveira_2006, Oliveira_2006, Cui_2008, Lourenco_2025, Samimi_2022}.

In this section, we review the main aspects of the generalized global entanglement, leading to the analysis derived in this work. The linear entropy is of practical use for assessing the presence of entanglement in a bipartite pure state. It is obtained as a first order approximation of the von Neumman entropy (Eq. (\ref{eq: von neumann entropy})) around the identity operator,  
\begin{equation}
    S_L (\rho_{\ell}) = \frac{d}{d-1} \left( 1 - \text{Tr} \rho_{\ell}^2 \right),
    \label{eq: linear entropy}
\end{equation}
where a normalization factor is introduced as to ensure that maximally entangled states satisfy $S_L(\rho_{\ell}) = 1$. The dimensional factor $d$ is the maximum of the dimension of subsystem $\ell$ and its complement, i.e., $d = \text{max} \left\{ d_{\ell}, d_{\bar{\ell}}\right\}$. 

For qubit systems, the trace of a reduced density matrix can be expressed in terms of expectation values of Pauli observables \cite{Fano1957}. Let $\rho$ be the density operator of a quantum system. Then, the one qubit purity, for the $i$-th qubit of the system, can be written as
\begin{equation}
    \text{Tr}(\rho_i^2) = \frac{1}{2} \sum_{\alpha = 0}^{3} \langle {\sigma_i^{\alpha}} \rangle ^2,
    \label{eq: purity one qubit}
\end{equation}
where $\sigma_i^{\alpha}$ are the Pauli matrices acting on the $i$-th qubit, with the identity being represented by $\sigma^0$. This expression can be readily generalized to subsystems composed of $n$ qubits \cite{Fano1957}: 
\begin{equation}
    \text{Tr}(\rho_{i_1, \cdots, i_n}^2) = \frac{1}{2^n} \sum_{\alpha_1, ..., \alpha_n = 0}^{3} \langle {\sigma_{i_1}^{\alpha_1} \otimes ... \otimes \sigma_{i_n}^{\alpha_n}} \rangle ^2.
    \label{eq: purity n qubit}
\end{equation}
This representation is essential for our study, as it enables entanglement assessment without requiring full reconstruction of the density matrix of the entire system. The notation $\rho_{i_1, ..., i_n}$ distinguishes individual qubits in a given partition of the system. For instance, setting $i_1 = 0$, $i_2 = 2$, and $i_3 = 4$, the notation $\rho_{024}$ represents the 3-qubit reduced density operator for the first, third, and fifth qubits.

For a bipartite system, the linear entropy of one partition already suffices to verify the existence of entanglement. However, to characterize multipartite entanglement, one needs to verify the entanglement between all possible partitions and their complements. Thus, the construction of $E_G^{(n)}$ as a quantifier of entanglement comes from considering all possible partitions of a given quantum system. In Ref.~\cite{Rigolin_2006}, the authors show that $E_G^{(n)}$ is given simply by the average of the linear entropy of $n$ qubit subsystems
\begin{equation}
    E_G^{(n)} = \llangle S_L( \rho_{i_1, ..., i_n} ) \rrangle,
    \label{eq: generalized entanglement}
\end{equation}
where the notation $\llangle \cdot \rrangle$ indicates the average over all possible ways of arranging such $n$ qubit subsystems. It is worth noting that this measure is indexed by $n$; therefore, it introduces different classes of entanglement. For instance, the second class builds upon the linear entropies of all possible arrangements of pairs of qubits. In turn, this allows for the proper quantification of entanglement, regardless of its distribution, in states where other measures fail (such as block entanglement \cite{Vidal_2003, Latorre_2004}). We restrict ourselves to the second and third classes, since in general, it suffices to fulfill this task for the studied states. However, higher classes may be needed for different states, depending on their internal structure~\cite{Rigolin_2006}. 

Although operationally straightforward, quantifying entanglement with $E_G^{(n)}$ requires the estimation of many operators that may grow substantially fast. From Eq.~\eqref{eq: purity n qubit}, the number of operators to be estimated for each partition scales as $4^n - 1$. Taking into account the average over all configurations from Eq.~\eqref{eq: generalized entanglement}, the total number of expectation values is given by
\begin{equation}
\label{eq: L operators}
    L = 
    \begin{pmatrix}
        N \\
        n
    \end{pmatrix} (4^n - 1),
\end{equation}
where $N$ is the number of qubits in the system, $n$ is the number of qubits per subsystem, and the binomial factor accounts for all possible partition configurations. If the system is invariant under particle permutation, such as W and GHZ states, $L$ can be considerably reduced, since all $n$-qubit partitions possess the same linear entropy. Then, one could, in principle, measure one partition and use the obtained linear entropy for all other possible partitions. However, here we also analyze tensor product states and random states. Note that in real quantum hardware, dissipation and decoherence can break particle-invariance symmetry because the qubits are not strictly identical.

Summarizing, this measure allows for the quantification of global entanglement without the need to explicitly reconstruct the density operator of the quantum state from experimental data. However, to do so, one needs to obtain precise estimates of several expectation values of physical observables, many of which are incompatible. Therefore, an appropriate method for this scenario is essential, as for larger systems, the measurement overhead can be a bottleneck. In the next section, we discuss examples of such methods, with particular emphasis on classical shadows.
\subsection{Classical shadows}
\label{subsec: classical shadows}

Efficiently extracting information from quantum systems is a fundamental step for numerous tasks in the present NISQ era, with limited resources and noisy devices~\cite{Preskill_2018}. In this scenario, classical shadows gained much attention, proving its value in a variety of applications \cite{Sack_2022, Vitale_2024, Rath_2021, Jerbi_2024, Abbas_2023, Garcia_2021, Faehrmann_2025} and has been experimentally demonstrated in different platforms, such as optical devices~\cite{Struchalin_2021, Zhang_2021, Liu_2022} and ion trap quantum computers~\cite{Stricker_2022}.

By formalizing a protocol that constructs an unbiased estimator of the density operator via randomized local measurements, the authors in Ref.~\cite{Huang_2020} demonstrated how classical shadows can be useful to estimate many expectation values from few measurements, with rigorous statistical guarantees, resulting in a sample complexity only logarithmic in the number of observables $L$ to be estimated. Since its original proposal, implementing shadow strategies for the estimation of multiple observables has been a valuable tool in previous works, highlighting its versatility in different applications, ranging from quantum chemistry to quantum algorithms~\cite{Huggins_2022, Blunt_2025, Basheer_2024, Benchen_Huang_2024, Shen_2024, Chan_2025, Boyd_2022, Ghisoni_2026, Bertuzzi_2025}. The capability of estimating many operators from few measurements is precisely the aspect of the protocol employed in this work.

Considering an unknown $ N$-qubit state, the protocol constructs a classical representation based on repetitions of a simple procedure: a unitary evolution that takes $\rho \rightarrow \rho' = U\rho U^{\dagger}$, followed by a~$Z$-basis measurement, which collapses the state to $|b\rangle \! \langle b|$, with $b \in \{ 0,1 \}^N$. For each repetition, the unitary $U$ is chosen at random from a fixed ensemble, and the collapsed state generates a classical snapshot $U^{\dagger} |b\rangle \! \langle b|U$. One can understand this process as the scrambling of the quantum information of the state to the measurement basis, given that each measurement of $\rho'$ in the computational basis is equivalent to a simultaneous measurement of $N$ commuting operators of the form $\left\{ U^{\dagger} Z_i U \right\}_{i = 1}^{N}$ on the original state~\cite{Hu_2022}.

Formally speaking, many repetitions of this procedure constitute a map that takes the unknown state to the average of the classical snapshots:
\begin{equation}
    \mathcal{M}(\rho) = \mathbb{E} \left( U^{\dagger} |b\rangle \! \langle b|U \right).
\end{equation}
Thus, the inversion of the map $\mathcal{M}$ enables one to reconstruct the density operator,
\begin{equation}
    \label{eq: inverse map}
    \rho = \mathbb{E} \left[ \mathcal{M}^{-1} \left( U^{\dagger} |b\rangle \! \langle b|U \right) \right] \equiv \mathbb{E} (\hat{\rho}),
\end{equation}
where we identify the estimator $\hat{\rho} = \mathcal{M}^{-1} \left( U^{\dagger} |b\rangle \! \langle b|U \right)$. Eq.~\eqref{eq: inverse map} states that $\hat{\rho}$ is an unbiased estimator, meaning its expectation value exactly recovers the original quantum state $\rho$. However, the empirical average converges to the true state only in the asymptotic limit of infinite measurements. In practice, reconstructing a state with such estimators using a finite sample can lead to non-physical states, whose density operators possess negative eigenvalues\cite{Huang_2020, Struchalin_2021}.

The ensemble of unitaries plays an important role in different variants of the original proposal of classical shadows. For instance, optimized versions of the protocol can be obtained considering generalized measurement settings~\cite{Nguyen_2022, Stricker_2022}, hamiltonian-driven unitaries~\cite{Hu_2022}, and hybrid setups which interplay local and global unitaries~\cite{Bertoni_2024}. In this work, we focus on local unitaries, as their implementation is suitable for current NISQ devices. When the unitaries are local, i.e. $U = \bigotimes_j U_j$, a typical choice is to sample $U_j$ from the ensemble of unitaries that rotate the computational basis to the $X$ and $Y$ basis: $U_j \in \{H, HS^{\dagger}, \mathbb{I} \}$. This characterizes the so-called Pauli measurements, which are particularly well suited for estimating Pauli words. In this case, the inversion of the quantum channel provides a closed-form expression for the estimators in Eq.~\eqref{eq: inverse map}. Here, we implement the multi-shot shadow estimation~\cite{Zhou_2023, Helsen_2023}: for each randomly selected measurement basis, $K$ shots are performed to achieve better convergence~\cite{Elben_2020}. Then, for the $m$-th measurement round, in possession of the measurement outcome bit string and the chosen unitaries, the estimators $\hat{\rho}$ read 
\begin{equation}
    \hat{\rho}^{(m)} = \frac{1}{K} \sum_{k = 1}^K \bigotimes_{j = 1}^N \left( 3U_j^{\dagger(m)} \big| b_j^{(m,k)} \big \rangle \! \big\langle b_j^{(m,k)} \big| U_j^{(m)} - \mathbb{I} \right),
    \label{eq: rho estimator}
\end{equation}
where $\mathbb{I}$ is the single qubit identity.

Considering a total of $M$ measurement schemes, the prediction step averages the estimated expectation value,
\begin{equation}
    \label{eq: expval estimator}
    \hat{o} = \frac{1}{M} \sum_{m = 1}^M \text{Tr} \left(O \hat{\rho}^{(m)} \right),
\end{equation}
where the observable $O$ is a Pauli string: $O = \bigotimes_j P_j$, with $P_j \in \{\sigma_x, \sigma_y, \sigma_z, \mathbb{I} \}$. It is worth noting that the original version of classical shadows considered single-shot settings, which is equivalent to setting $K=1$ in Eq.~\eqref{eq: rho estimator}. Then, combining Eqs.~\eqref{eq: rho estimator} and~\eqref{eq: expval estimator}, we can rewrite the expectation value of a Pauli string as
\begin{equation}
    \hat{o}  = \frac{1}{MK} \sum_{m,k} \prod_{j=1}^{n} \text{Tr} \biggl( \frac{3}{2} (1 - 2b_j^{(m,k)}) P_j \sigma_{j,(m)} + \frac{1}{2}P_j \biggr), 
    \label{eq: expval Paulis}
\end{equation}
where $\sigma_{j,(m)}$ is the Pauli matrix associated with the unitary  $U_j^{(m)}$, in which the action of the unitary on the computational basis returns the eigenstates of $\sigma_{j,(m)}$, and $ b_j^{(m,k)} \in \{0,1\} $ are computational basis outcomes. Due to the orthogonality of Pauli matrices, the trace term yields only three possible values:
\begin{equation}
\label{eq: matching condition}
    \begin{cases}
        1 & \text{if } P_j = \mathbb{I}, \\
        0 & \text{if } P_j \neq \sigma_{j,(m)}, \\
        \pm 3 & \text{if } P_j = \sigma_{j,(m)}.
    \end{cases}
\end{equation}
Only when each $ P_j $ matches the measurement basis $ \sigma_{j,(m)} $ does the estimator return a nonzero contribution. Hence, the final observable estimate is built from the subset of measurement rounds where all such matches occur:
\begin{equation}
    \hat{o} \approx \frac{1}{\widetilde{M}K} \sum_{\tilde{m},k} \prod_{j} (1 - 2b_j^{(\tilde{m},k)}),
\end{equation}
where $ \widetilde{M} $ is the number of matched rounds.
Thus, when calculating the expectation value of Pauli observables, one needs simply to check the correspondence between the randomly chosen bases and the Pauli matrices composing the observable. In the language of Ref.~\cite{Huang_2021}, we say that a measurement scheme ``hits'' the observable when the single qubit unitary matches the Pauli operator in $O$. If the chosen basis hits the observable, we accumulate the eigenvalue of the collapsed state $|b\rangle$. If the operator in $O$ is the identity, then nothing happens. However, if there is a mismatch, the trace in Eq.~\eqref{eq: expval estimator} is equal to zero, and we get no information about the expected value with such measurement scheme. As a result, classical shadows equipped with Pauli measurements provide a simple and efficient way for the estimation of Pauli observables.

The sample complexity for estimating $L$ operators is given by~\cite{Elben_2022}
\begin{equation}
    M = \mathcal{O} \left( \frac{3^w \log(L)}{\epsilon^2} \right),
    \label{eq: sample complexity bound}
\end{equation}
where $w$ is the maximum locality of the $L$ operators and $\epsilon$ is the the additive error bound for each observable: $|\text{Tr}(O \rho) - \hat{o}|\leq \epsilon$. In spite of the exponential factor in the sample complexity, if one is interested in properties of small subsystems or operators with low locality, classical shadows stands as a favorable approach. This is particularly well suited for the analysis of the generalized global entanglement measure $E_G^{(n)}$, given that $w$ is fixed and small compared to the number of qubits. Additionally, even though the number of observables to be estimated scales rapidly, the logarithmic dependence in Eq.~\eqref{eq: sample complexity bound} guarantees it is not an overhead. 

The matching condition in Eq.~\eqref{eq: matching condition} is a possible drawback regarding randomized shadows, since unmatched measurement rounds are discarded. When estimating a Pauli observable, the probability of hitting any of its Pauli matrices is $1/3$, so the probability for a complete hit is $1/3^{w}$. This is a more intuitive way to understand the factor $3^w$ in Eq.~\eqref{eq: sample complexity bound}. To circumvent this probabilistic issue, a derandomized version of classical shadows was developed~\cite{Huang_2021}, tailored for estimating Pauli strings typically present in molecular Hamiltonians. This protocol chooses the measurement basis, allocating the measurement budget with respect to the weight of the operators in a given Hamiltonian. This, in turn, assures that there is always a match between the measurement basis and (at least) one of the observables, therefore avoiding the discard of any measurement data. Despite improving the accuracy of the estimation, this procedure introduces a preprocessing step that can be expensive and scales rapidly with the number of measured operators.
  
\subsection{Direct estimation methods}
\label{subsec: direct estimation}

The task of estimating multiple observables efficiently can also be tackled with direct estimation methods, i.e., approaches that require minimal post-processing of the experimental data. One could, in principle, estimate each observable individually; however, such a procedure is not optimized in any sense. A possible means to optimize the use of measurement resources is to group compatible observables, allowing the estimation of multiple operators from the same measurement data by rotating the system to a basis in which those operators are diagonal. This can be especially powerful when considering low locality operators acting on multi-qubit systems. Nonetheless, this grouping procedure is not trivial, and several approaches have been studied for this purpose. 

A common approach is to group observables considering qubit-wise commutativity (QWC). In general, this method maps the grouping problem to a graph theory problem. One can construct a graph, mapping each operator to a node, with edges connecting qubit-wise commuting operators. The grouping choice is equivalent to the minimum clique cover: partitioning the whole graph into complete subgraphs, such that the number of subgraphs is the minimum possible. Then, each subgraph represents a group of compatible observables, and the minimal number of groups assures an optimized measurement scheme~\cite{Verteletskyi_2020_qwc}. The circuit to measure the compatible observables can then be realized with single-qubit unitaries. Since this problem is known to be NP-hard in graph theory, such grouping relies on preprocessing steps that can be computationally expensive, particularly as the number of qubits and complexity of the problem grow. Another possibility is to consider full commutativity (FC) between operators. This condition is weaker than requiring QWC, however, much simpler in terms of preprocessing steps. In spite of being computationally favorable, this method has practical complications given that it requires larger circuits, composed of entangling gates~\cite{Crawford_2021_efficientquantum}. 

In order to test the feasibility of the shadow approach, we evaluated the global entanglement measure for some cases of interest, comparing shadow estimations with QWC and FC estimations. 
\section{Results and discussion}
\label{sec: results}
The original shadow estimation protocol relies on a single-shot procedure, in which a new random basis is chosen for every individual measurement~\cite{Huang_2020}. To simplify hardware implementation, a more practical approach involves applying fewer basis rotations and collecting multiple shots per basis, provided the statistical accuracy of the estimations is maintained. Thus, we explore scenarios with different distributions of random measurement schemes and shots per scheme. We define the ratio $r = M/K$, in a similar way as done in Ref.~\cite{Brydges_2019}, between the total number of random measurement schemes and the number of shots per scheme. Then, we and compare different measurement procedures, all with the same measurement budget, $T = MK$. So, we take $T$ and $r$ as inputs, set $M = \sqrt{rT}$ and $K = \sqrt{T/r}$, and take the ceiling of $M$ and $K$ for numerical implementations.

For the QWC and FC estimations, we distribute the total number of measurements uniformly with respect to the number of groups of compatible observables, denoted by $C$. Then, each group has a measurement budget of $MK/C$. To test different experimental scenarios, we fixed $r \in \{0.5, 1.0, 5.0 \}$, to represent cases with: fewer repetitions but many random measurement bases ($r = 5$), fewer measurement bases but many repetitions ($r = 0.5$), and an intermediate scenario ($r = 1$). Then, varying the total number of measurements, we ran our simulations using Ket Platform~\cite{Rosa_2026}, considering 10-qubit GHZ, W, EPR, and random states, as follows:
\begin{equation}
  \begin{aligned}
    &|\psi_{\text{GHZ}}(N) \rangle = \frac{1}{\sqrt{2}} \left( |0\rangle^{\otimes N} + |1\rangle^{\otimes N} \right), \\
    &|\psi_\text{W}(N) \rangle = \frac{1}{N} \sum_{j=1}^{N} |00 \cdots 1_j \cdots 00 \rangle, \\
    &|\psi_{\text{EPR}}(N) \rangle = |\Phi^{+}\rangle^{\otimes N/2}\ \\
    &|\psi_{\text{rand}}(N) \rangle = U_\text{Haar} |0\rangle^{\otimes N},
  \end{aligned}
\end{equation}
where $|\Phi^{+}\rangle^{\otimes N/2}$ stands for the usual 2-qubit EPR state~\cite{Bell_1964} and $U_\text{Haar}$ is a unitary operator randomly sampled with respect to the Haar measure~\cite{Mezzadri_2007}. 

\begin{figure}
    \centering
    \includegraphics[width=\linewidth]{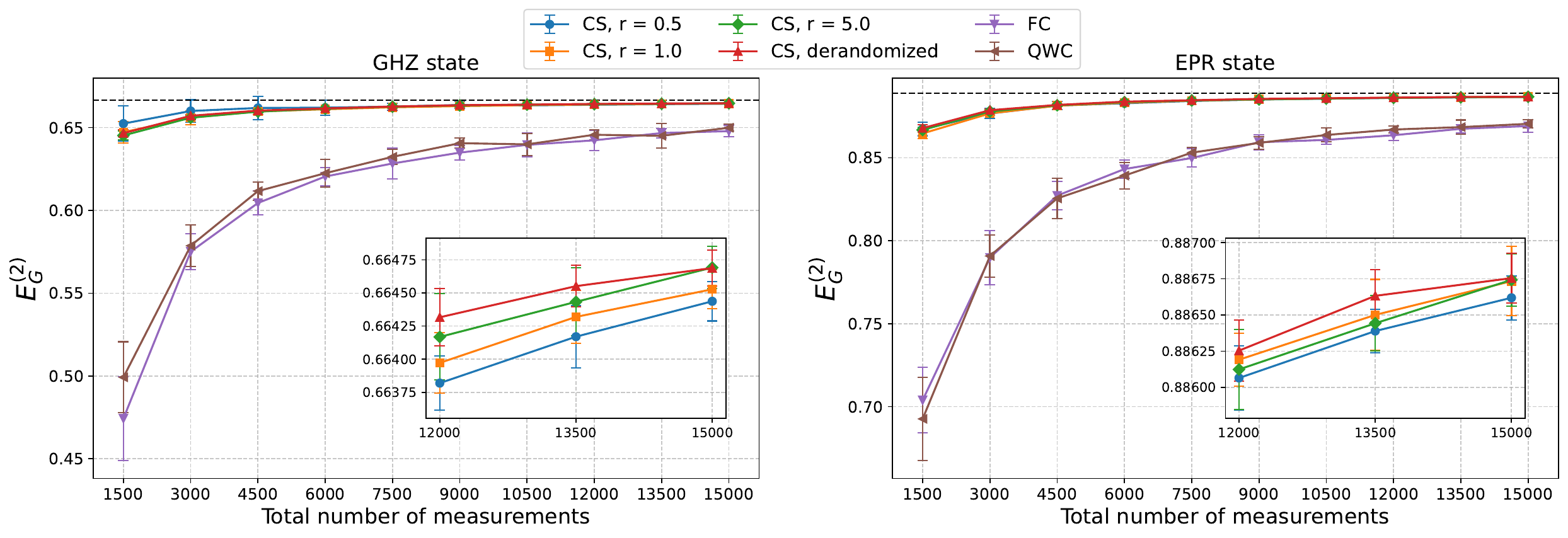} \\
    \includegraphics[width=\linewidth]{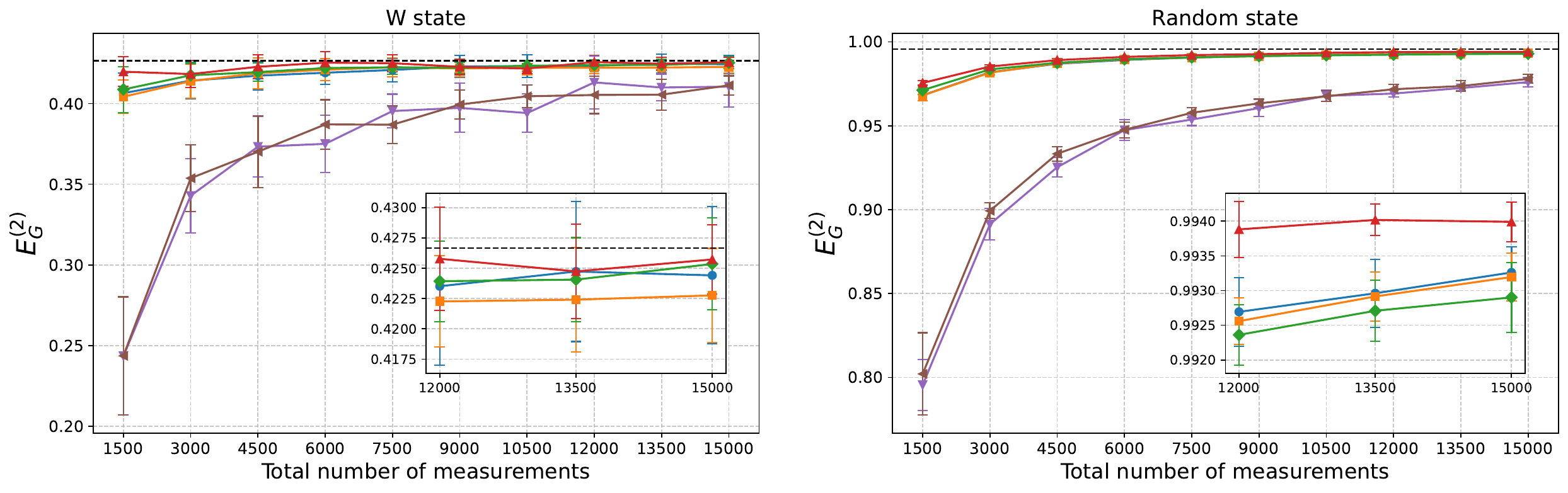} \\
    \caption{$E_G^{(2)}$ as a function of the total number of measurements for 10-qubit GHZ, EPR, W, and random states, requiring the estimation of 675 Pauli words. The results show a faster convergence to the exact entanglement value (dotted line) for shadow estimations, and the insets highlight how, even with different values of $r$, the randomized shadow estimations perform as well as the derandomized shadow protocol. The uncertainty bars represent the standard deviation over ten independent runs.}
    \label{fig: EG_2}
\end{figure}

\begin{figure}
    \centering
    \includegraphics[width=\linewidth]{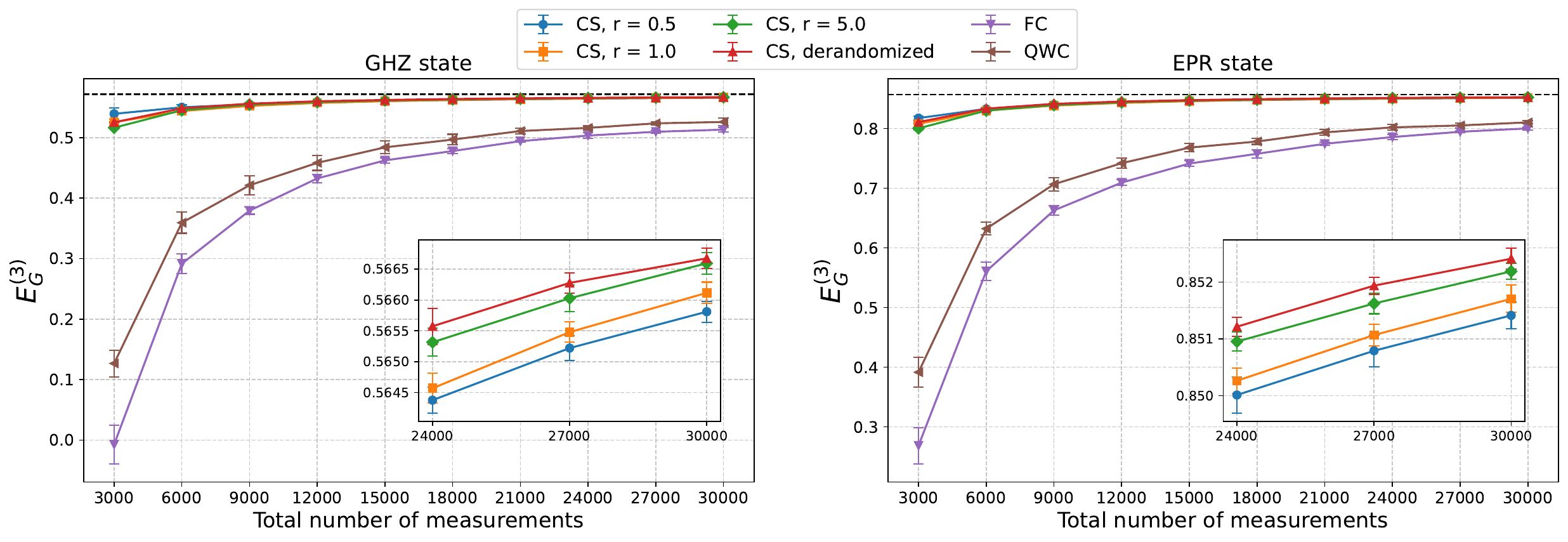} \\
    \includegraphics[width=\linewidth]{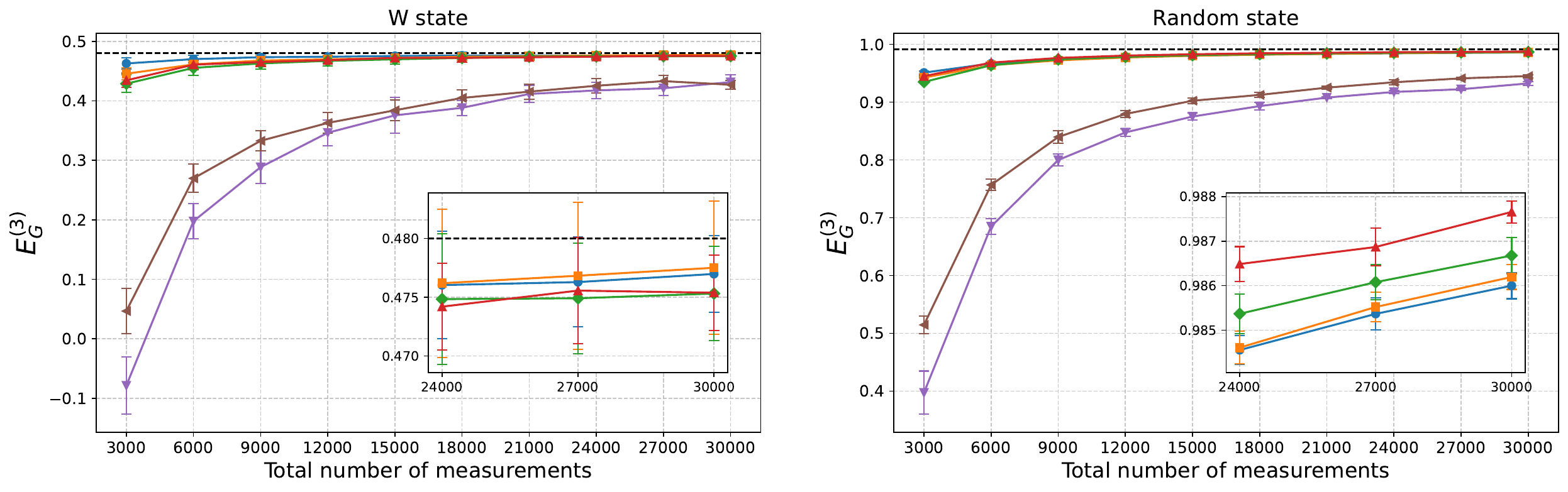}
    \caption{$E_G^{(3)}$ as a function of the total number of measurements for 10-qubit GHZ, EPR, W, and random states. The results strengthen the observations in Fig~\ref{fig: EG_2}, tackling a significantly harder estimation task, involving 7560 observables to characterize $E_G^{(3)}$. The uncertainty bars represent the standard deviation over ten independent runs.}
    \label{fig: EG_3}
\end{figure}

The core of evaluating the generalized global entanglement lies in the estimation of the linear entropies of all possible $n$-qubit partitions of a given state (see Eq.~\eqref{eq: generalized entanglement}; ultimately, this reduces to the estimation of many expectation values, given by Eq.~\eqref{eq: purity n qubit}. Fixing a maximum measurement budget of  $T = 15 \times 10^3$ measurements, we evaluate $E_G^{(2)}$ and $E_G^{(3)}$ for the aforementioned states. The results presented in Fig.~\ref{fig: EG_2}, for the estimation of $E_G^{(2)}$, stress that classical shadows offers a clear advantage over QWC and FC estimations in the estimation of the global entanglement between two qubits and the rest of the system, converging to the exact results with much fewer measurements. Fig~\ref{fig: EG_3} further stresses that advantage, as the difference of the shadow methods versus direct estimation methods are larger, in a more challenging scenario: in order to estimate $E_G^{(3)}$ for 10-qubit states, 7560 expected values are needed, as opposed to 675 operators involved in the estimation of $E_G^{(2)}$.

\bgroup
\def\arraystretch{1.15}
\begin{table}
    \centering
    \caption{Relative percentage error $\Delta_{\%}$ and standard deviation for the second and third classes of $E_G^{(n)}$ considering a 10-qubit GHZ state. The data is taken from Fig.~\ref{fig: EG_2} and Fig.~\ref{fig: EG_3}, with $6 \times 10^3$ and $30 \times 10^3$ measurements, respectively. }
    \begin{ruledtabular}
        \begin{tabular}{cccccc}
        \makecell{Entanglement \\ class}    & Method   & $M$    & $K$   & $\Delta_{\%}$ & \makecell{Standard \\ deviation} \\
        \colrule
        \multirow{6}{*}{$E_G^{(2)}$}    & CS $(r = 0.5)$    & 55    & 110   & 0.6920 & 0.0045 \\
        &   CS $(r = 1.0)$  & 78        & 78                & 0.8384 & 0.0004  \\
        &   CS $(r = 5.0)$  & 174       & 35                & 0.7830 & 0.0006  \\
        &   CS (derand)     & -         & -                 & 0.7610 & 0.0005  \\
        &   QWC             & -         & -                 & 6.6305 & 0.0085  \\
        &   FC              & -         & -                 & 6.9288 & 0.0055  \\
        \colrule
        \multirow{6}{*}{$E_G^{(3)}$}    & CS $(r = 0.5)$    & 123  &  245  &  0.9831 & 0.0002 \\
        &   CS $(r = 1.0)$  & 174       & 174               &  0.9297       & 0.0002  \\
        &   CS $(r = 5.0)$  & 388       & 78                &  0.8460   & 0.0002  \\
        &   CS (derand)     & -         & -                 &  0.8319   & 0.0002  \\
        &   QWC             & -         & -                 &  7.9853   & 0.0061  \\
        &   FC              & -         & -                 &  10.2155  & 0.0038  \\
        \end{tabular}
    \end{ruledtabular}
    \label{tab: table relative error}
\end{table}
\egroup
 Table~\ref{tab: table relative error} displays the relative error in the estimation of $E_G^{(2)}$ and $E_G^{(3)}$ for a 10-qubit GHZ state with a fixed measurement budget. Specifically, the data is taken from  Fig.~\ref{fig: EG_2} and Fig.~\ref{fig: EG_3}, and indicate when the shadow estimations reach a relative error of less than one percent. Notably, the shadow estimations with $r = 0.5$ indicate that the global entanglement can be accurately estimated even if the number of experimental settings is limited. In line with this, it is worth discussing multi-shot shadows in more detail. The expression of the variance for estimating a $w$-local Pauli observable (Theorem 2 in Ref.~\cite{Zhou_2023}) is
\begin{equation}
    \text{Var}(\hat{o}) = \frac{1}{M} \left[ \frac{3^w}{K} + \left( 1 - \frac{1}{K}\right)3^w \text{Tr} (O\rho)^2 - \text{Tr} (O\rho)^2  \right], 
    \label{eq: variance multi-shadows}
\end{equation}
which, for the extremal cases of $\text{Tr} (O\rho) = 1$ and $\text{Tr} (O\rho) = 0$, gives $\text{Var}(\hat{o})=(3^w - 1)/M$ and $\text{Var}(\hat{o})~=~(3^w)/MK$, respectively. In the first case, $K$ does not affect the variance, so additional repetitions provide no benefit. In the latter, the variance scales inversely with the total number of measurements, $MK$, allowing fewer measurement settings with more repetitions. In the intermediate case, the full expression for the variance can be used to obtain an optimal allocation of measurement resources. For a fixed total number of measurements $T=MK$, it is easy to check that Eq.~\eqref{eq: variance multi-shadows} has a minimum when $K=1$. Thus, adding repetitions at the cost of reducing the number of random measurement settings would not be the best choice. Yet, for this particular entanglement measure, the large collection of observables required leads to interesting properties. For GHZ, W and EPR states, only a few expected values required to estimate $E_G^{(n)}$ are different than zero, profiting from the second case discussed -- however, this information is only available if one already knows the state being measured. In a more general scenario, considering a Haar-random state, most of the expected values are sufficiently small, so the dominant term in the variance still depends on the total number of measurements, $MK$. To summarize, the choice of how to distribute the measurement budget certainly affects the accuracy of the estimation, but $E_G^{(n)}$ allows for a flexibility in this choice, rendering simpler experimental procedures, while keeping accurate estimates.
\begin{figure}
    \centering
    \includegraphics[width=\linewidth]{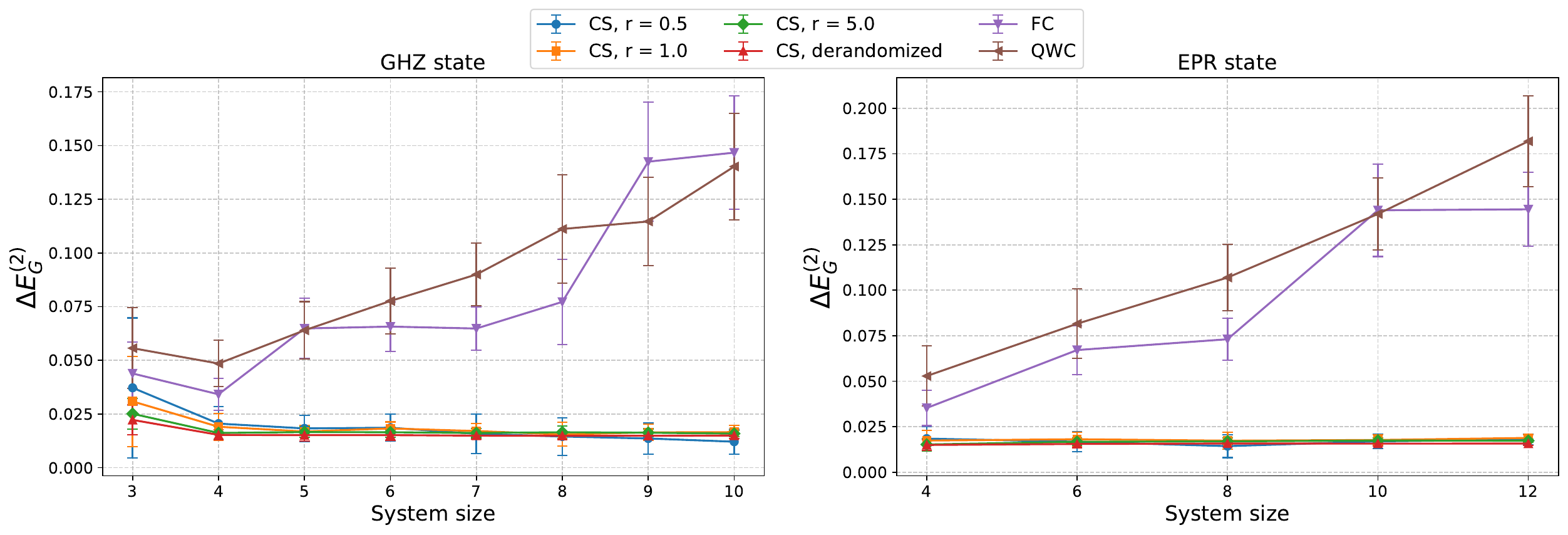} \\
    \includegraphics[width=\linewidth]{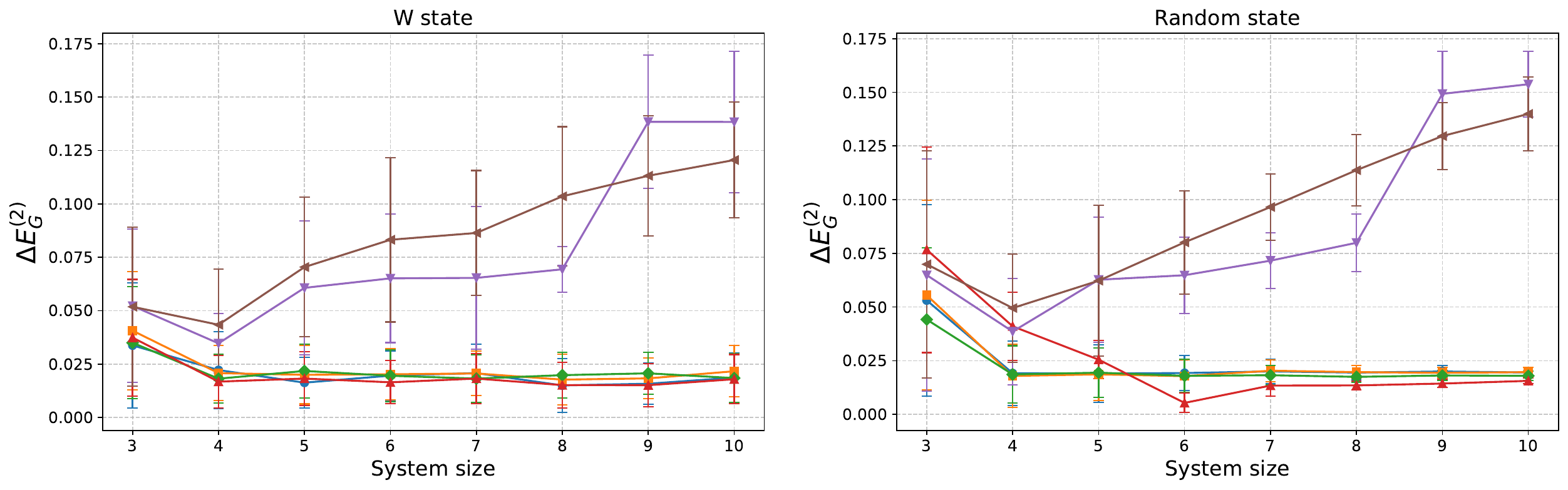} \\
    \caption{Absolute difference between the exact and estimated values of $E_G^{(2)}$ for different system sizes, considering GHZ, EPR, W, and Haar random states. When the system size grows, the absolute error estimated with grouping methods also grows, while the shadow approach is able to maintain the errors relatively low.}
    \label{fig: EG_2 diff}
\end{figure}
Additionally, the performance of the randomized shadow estimation was practically the same as the derandomized one, emphasizing that, for this particular entanglement measure, one can obtain good results with fewer measurements without resorting to expensive preprocessing steps. This equivalence in performance is directly linked to how the generalized global entanglement measure is constructed. As we need to average the linear entropies of all possible $n$-qubit partitions in a given system, all possible $n$-qubit Pauli words have to be estimated. This situation is the opposite of the cases where derandomized shadows thrive, typically when dealing with specific weighted operators in molecular Hamiltonians~\cite{Huang_2021}. Therefore, the case for $E_G^{(n)}$ stands as an accessible means to quantify entanglement using simple measurement procedures. 

Not only do shadow estimations converge faster to the exact value of $E_G^{(2)}$, but the difference in performance is even greater in a case with very few measurements -- as can be seen for the first few points in Fig.~\ref{fig: EG_2} and in Fig.~\ref{fig: EG_3}. In light of that, we considered a very restricted scenario,  with a measurement budget of only $MK = 2 \times 10^3$ measurements, and studied the estimation of $E_G^{(2)}$ for the same states, increasing the number of qubits in the system and computing the absolute difference between our estimations and their exact counterparts, which we denoted by $\Delta E_G^{(2)}$. The results displayed in Fig.~\ref{fig: EG_2 diff} show that, even in this very limited scenario, the error estimated with classical shadows remained relatively small and well-behaved as the system size grows, while the error for grouping estimations grows with the size of the system. Increasing the number of qubits in the state, we require the evaluation of more observables, making the estimation of $E_G^{(2)}$ a harder task with such a limited measurement budget. Hence, for grouping strategies, the number of measurements per group decreases as the number of qubits increases, leading to a less precise estimation. This is not the case for shadow estimation: because the estimator in Eq.~\eqref{eq: rho estimator} is a tensor product structure, classical shadows can predict more observables as the system size increases.
\section{Conclusion}
\label{sec: conclusion}
We studied the quantification of entanglement via $E_G^{(n)}$, a measure introduced in Ref.~\cite{Rigolin_2006}. The core of performing such a task lies in the estimation of many Pauli words, a number that can grow rapidly as systems with more qubits are considered. We simulated a more realistic scenario in which the measurement budget is limited, comparing grouping methods, such as full commutativity and qubit-wise commutativity, with randomized and derandomized procedures of classical shadows. The analysis made with well-known entangled states (GHZ, W, tensor product of EPR states) and random states indicates that classical shadows provides a faster convergence to the exact value of entanglement (Fig.~\ref{fig: EG_2}). Also, as the number of Pauli words grows for larger systems, the difference in performance increases in favor of classical shadows, and a significantly better performance for shadow estimations is observed in regimes where the number of measurements is very limited, which is depicted in Fig.~\ref{fig: EG_2 diff}. Our results suggest that, in cases where the precise estimation of a large number of observables is required, random measurement schemes can make such tasks feasible in scenarios with limited resources.
\begin{acknowledgments}
The authors acknowledge the financial support provided by the Brazilian funding agencies Coordenação de Aperfeiçoamento de Pessoal de Nível Superior (CAPES) and Conselho Nacional de Desenvolvimento Científico e Tecnológico (CNPq) through Grant No. 409673/2022-6. EID and JPE also acknowledge the support of the Instituto Nacional de Ciência e Tecnologia de Informação Quântica (INCT-IQ) under Grant No. 409673/2022-6, Instituto Nacional de Ciência e Tecnologia de Infraestruturas Quânticas e Nano para Aplicaç\~oes Convergentes INCT-IQNano under Grant No. 406636/2022-2, and Instituto Nacional de Ciência e Tecnologia de Computação Quântica Aplicada under Grant No. INCT-CQA 408884/2024-0.
\end{acknowledgments}

\bibliography{apssamp}

\end{document}